**Sperry versus Hebb: Topographic mapping in Isl2/EphA3 mutant mice.**

*Dmitry Tsigankov and Alexei Koulakov*

*Cold Spring Harbor Laboratory, Cold Spring Harbor, NY 11724*

In wild-type mice axons of retinal ganglion cells establish topographically precise projection to the superior colliculus of the midbrain. This implies that axons of neighboring retinal ganglion cells project to the proximal locations in the target. The precision of topographic projection is a result of combined effects of molecular labels, such as Eph receptors and ephrins, and correlated electric activity. In the Isl2/EphA3 mutant mice the expression levels of molecular labels is changed. As a result the topographic projection is rewired so that the neighborhood relationships between retinal cell axons are disrupted. Here we argue that the effects of correlated activity presenting themselves in the form of Hebbian learning rules can facilitate the restoration of the topographic connectivity even when the molecular labels carry conflicting instructions. This occurs because the correlations in electric activity carry information about retinal cells' spatial location that is independent on molecular labels. We argue therefore that experiments in Isl2/EphA3 knock-in mice directly test the interaction between effects of molecular labels and correlated activity during the development of neural connectivity.

## Introduction

In developing brain connectivity is established under the influence of several factors. Neurons initially find appropriate targets based on the sets of chemical labels (Sperry, 1963; Tessier-Lavigne and Goodman, 1996; O'Leary et al., 1999). This *chemospecificity hypothesis* originally postulated by Rodger Sperry (Sperry, 1963) motivated the search for molecules that could be used as such cues and suggested the principles which direct growing axons to their targets. The precision of axonal projections is further fine-tuned through mechanisms based on correlated neural activity. These activity-dependent mechanisms are thought to implement the rules for modification of neuronal connections that were proposed by Donald Hebb (Hebb, 1949). Hebbian rules provide a paradigm through which sensory experience may influence the formation of the neuronal connectivity. This is in contrast to the molecular labels that are controlled primarily by genes. One of the central questions in the studies of the developing nervous system is how the influences of molecular labels are combined with Hebbian learning rules to yield connectivity that is both precise and adaptive (Cline, 2003).

The interaction between molecular cues and activity-dependent factors has been extensively studied on the example of the topographic projection from retina to superior colliculus (SC) (Cline, 2003). Axons of retinal ganglion cells (RGC) form an orderly representation of the visual world in the brain, called topographic or retinotopic map (Kaas, 1997). This implies that two RGC axons, which originate from neighboring points in retina, terminate next to each other in the target region. Topographic maps are important to the organism, because they facilitate visual processing, which involves wiring local to the termination zone (Cowey, 1979; Chklovskii and Koulakov, 2004).

In case of retintotopic mapping the role of molecular labels is played by the Eph family of receptor tyrosine kinases and their ligands ephrins (Drescher, 1997; Feldheim et al., 1998; Flanagan and Vanderhaeghen, 1998; Goodhill and Richards, 1999; O'Leary and Wilkinson, 1999; Feldheim et al., 2000; McLaughlin et al., 2003a; Pasquale, 2005). The coordinate system is encoded in the retina through graded expression of Eph receptors by RGCs (Cheng and Flanagan, 1994; Holash and Pasquale, 1995; Flenniken et al., 1996; Marcus et al., 1996; Zhang et al., 1996; Birgbauer et al., 2000; Hindges et al., 2002; Mann et al., 2002). The recipient coordinates in SC are established by the graded expression of ephrin ligands (Braisted et al., 1997; Frisen et al., 1998; Feldheim et al., 2000; Hindges et al., 2002). The layout of the map in Figure 1 implies that axons expressing high level of EphA receptor are repelled by areas in the target expressing high level of ephrin-A. Similarly it follows that the axons with high concentration of EphB label are attracted to the ephrin-B-rich regions (see however (McLaughlin et al., 2003c; Tsigankov and Koulakov, 2004) for a discussion of possible alternatives). It is also possible that other molecules are involved in the formation of this projection such as neurotrophins (Marotte et al., 2004).

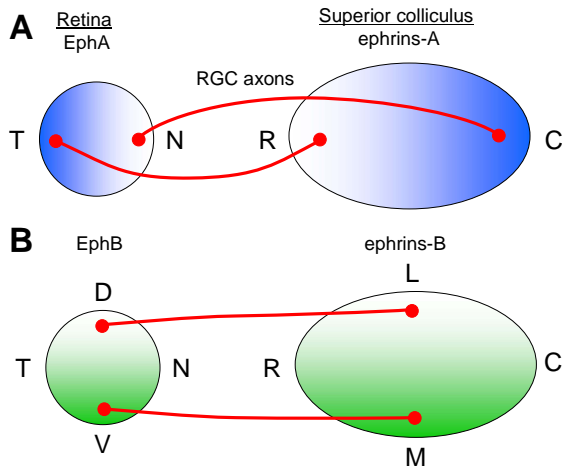

**Figure 1.** Topographic map from retina to superior colliculus is dependent on two systems of reciprocal gradients. (**A**) Temporal-nasal (TN) axis of the retina is mapped onto the anterior-posterior (AP) [also called rostral-caudal (RC)] axis in superior colliculus. This mapping is established through the graded expression of EphA receptors in retina and reciprocal expression of ligands from ephrin-A family in SC. (**B**) Similar principle applies to the mapping of dorsal-ventral (DV) axis of the retina onto medial-lateral (ML) axis of the target, where topography is formed based on gradients of EphB/ephrin-B pair.

The precision of topographic projections is further enhanced through mechanisms based on correlated neural activity. Due to the presence of retinal waves during critical period of development, electric activity is similar in RGC axons neighboring in retina (McLaughlin et al., 2003b; Pfeiffenberger et al., 2005). Correlated activity therefore provides additional information about axonal geometric origin in retina. Topographic maps are disrupted by blockade of the afferent activity with TTX (Thompson and Holt, 1989; Constantine-Paton et al., 1990), block of NMDA receptor in the target (Cline and Constantine-Paton, 1989; Schmidt, 1990), or disruption of retinal waves during development (McLaughlin et al., 2003b; Chandrasekaran et al., 2005; Mrsic-Flogel et al., 2005; Pfeiffenberger et al., 2005). The rough resolution of topographic maps is however preserved after these manipulations (Schmidt, 1990; McLaughlin et al., 2003b) due to remaining chemical labels. Therefore patterned electric activity contributes to the refinement of topographic projection while Ephs and ephrins determine the global ordering of projections (Cline, 2003; McLaughlin and O'Leary, 2005; Tsigankov and Koulakov, 2006).

Insights into the mechanisms of map formation can be obtained from the experiments with mutant mice in which the distribution of chemical cues is altered. Thus, in Isl2/EphA3 knock-in mice a randomly chosen subset of retinal cells expresses additional EphA3 receptor that is not found in the RGCs of wild-type retina (Brown et al., 2000). As a result the connectivity between retina and SC becomes changed. Simple models based on axonal sorting on the basis of the overall level of EphA expression succeed in explaining most of the phenotypes observed in Isl2/EphA3 mutants (Koulakov and Tsigankov, 2004; Reber et al., 2004). In some cases however the connectivity fails to change in these animals despite substantial modification of chemical labels (Brown et al., 2000). Here we argue that the robustness of connectivity with respect to genetic manipulations may stem from the correlated activity-based Hebbian rules that remain operational in mutant retina. We suggest therefore that experience-dependent contribution arising from Hebbian rules may negate the effects of chemical labels. Our study suggests that experiments in Isl2/EphA3 mutant mice directly test the interplay between effects of molecular labels and correlated neural activity.

## Results

### Topographic connectivity in wild type animals

The mechanisms for the formation of topographic maps have received an extensive attention from theorists (Honda, 2003; Reber et al., 2004; Yates et al., 2004; Goodhill and Xu, 2005; Tsigankov and Koulakov, 2006). The simplest model that involves Eph-ephrin signaling and competition between axons was recently developed by us (Koulakov and Tsigankov, 2004; Tsigankov and Koulakov, 2006). The model includes repulsive interactions between axons expressing EphA receptors and the dendrites in SC that express ephrin-A ligands. The model postulates that the retino-collicular connectivity attempts to minimize the total number of EphA receptors bound to ligands evaluated for the entire system of axons. This view is consistent with the systems-matching ideas derived from the early lesion and transplantation experiments (Yoon, 1975, 1976). The total number of bound/activated receptors is calculated using a simple form of mass action law (see Methods for more detail). It is then assumed that during retino-collicular development the connectivity evolves to minimize the total number of bound receptors thorough the iterative process of axon remodeling.

If the interactions in EphA/ephrin-A receptor/ligand pair were the only factor, all axons would project to the locations in the target with the lowest level of ligand (rostral in Figure 1). To prevent this it is assumed that axons compete for space or limiting factors in the target. Competition forces the axons with low levels of receptor to terminate in the areas with higher levels of repellent ligand, since these axons are more indifferent to the ligand. Thus axonal competition in combination with EphA/ephrin-A signaling leads to the formation of the ordered topographic projections along AP axis. The importance of competition in the development of topographic connectivity was first emphasized by Prestige and Willshaw (Prestige and Willshaw, 1975). These authors also predicted the accurate distribution of repellent molecular labels that was later confirmed with the experimental discovery of Ephs and ephrins (Flanagan and Vanderhaeghen, 1998). Since then competition has emerged as an important factor in many genetic and surgical experiments (Yoon, 1977; Feldheim et al., 2000). Similarly, the competition and chemoattraction in EphB/ephrin-B pair leads to the topographic mapping along ML axis if the entire system of axons tends to maximize the total number of EphB receptor bound to the ligand.

Our approach can also account for the effects of correlated activity on the topographic connectivity. These effects are governed by the Hebbian plasticity rules and can be included in the model using similar approach as with the molecular labels. When accounting for molecular labels, we suggested that topographic connectivity minimizes the total number of receptors bound/activated by ligands. This number is described in our model by quantity $E_{chem}$ that is defined more precisely in the Methods. Thus, the effects of molecular labels in our model are accounted for by minimizing the value of $E_{chem}$. Similarly the effects of correlated activity are included by minimizing another quantity $E_{act}$ (see Methods). Two contributions $E_{chem}$ and $E_{act}$ are combined in our model additively

$$E = E_{chem} + E_{act} \qquad (1)$$

To describe the developing system of retinocollicular projections we minimize the sum described by equation (1) computationally or analytically (using pencil and paper, see Appendix A). The minimization approach postulated in equation (1) was pioneered by Fraser and Perkel before Eph receptors and ephrin ligands were known as the topographic labels (Fraser and Perkel, 1990). We adopted this approach recently to describe Eph/ephrin-based signaling (Koulakov and Tsigankov, 2004). We also argued that additive form of interaction between molecular labels and the effects of correlated activity postulated in equation (1) is consistent with the experiments in ephrin-A knockout mice (Tsigankov and Koulakov, 2006).

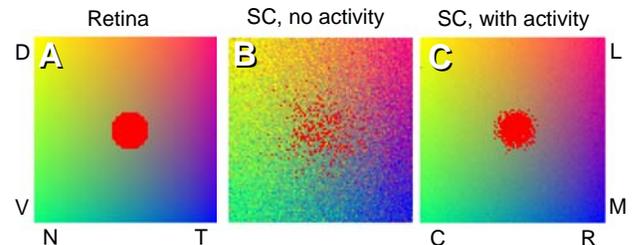

**Figure 2.** Hebbian rules lead to an effective attraction in the target between axons with correlated activity. (A) The origin of axons in the retina is color-coded. A small subset of axons in the area representing the middle of the retina is selected for tracing (red). (B) In the absence of correlated activity $E_{act} = 0$ the topographic projection is imprecise and the set of axons selected in (A) forms a diffuse cloud in the target. (C) When $E_{act} \neq 0$ the axons neighboring in retina (red) carry correlated activity and are attracted to each other. This effective attraction leads to the condensation of the axons into an almost precise image of the circle in the retina shown in (A). Thus Hebbian contribution $E_{act}$ leads to the sharpening of topographic projection.

For our subsequent discussion it is important to understand the effect of correlated activity-dependent contribution $E_{act}$. This effect was derived recently from the Hebbian learning rules (Tsigankov and Koulakov, 2006). We argue that Hebbian rules lead to an effective pair-wise attraction in the target between axons with correlated activity. This attraction makes topographic projection more precise (Figure 2). Sharpening of topographic projection due to the effects of correlated activity is consistent with experimental data on partial or full activity blockade (McLaughlin et al., 2003b).

In wild-type animals chemospecificity in the form of Eph/ephrin-based signaling and the Hebbian rules operate in unison leading to essentially the same ordering of axons (Figure 2). Mathematically this implies that minimum of $E_{chem}$ coincides with that of $E_{act}$. The goal of both of these factors is to cooperate in making the topographic projection as sharp as possible. Below we will consider the connectivity in Isl2/EphA3 mutant mice for which the cooperation between chemical factors and Hebbian rules is disrupted.

**Retinocollicular connectivity in Isl2/EphA3 knock-in mice.**

The distribution of EphA receptors is altered in retinas of Isl2/EphA3 knock-in mice (Brown et al., 2000; Honda, 2003; Koulakov and Tsigankov, 2004; Reber et al., 2004). In these animals roughly 50% of retinal cells receive an increase in the level of receptor EphA3, which is not expressed by the wild-type cells. Thus, at each position in the retina there are two classes of cells: EphA3 positive and negative (EphA3+ and EphA3-, Figure 3A). This is accomplished by coexpression of ectopic EphA3 with LIM homeobox transcription factor Islet2 (Isl2) that is found in mosaic binary pattern of expression throughout RGC layer. EphA3+ axons experience a stronger repulsion from ligand in the target (SC) than EphA3- axons. Therefore EphA3+ and EphA3- axons neighboring in retina should terminate at different positions in the target. In particular the EphA3+ cells are expected to terminate at positions with lower level of repellent (Figure 1), i.e. at more rostral positions than EphA3- axons. This consideration leads to the prediction that tracing of axons projecting from a single point in retina should yield two TZs in SC. One TZ (rostral) corresponds to EphA3+ axons, while the caudal TZ corresponds to EphA3- axons. This prediction is indeed confirmed experimentally in homozygous knock-ins (Brown et al., 2000) (Figure 3B).

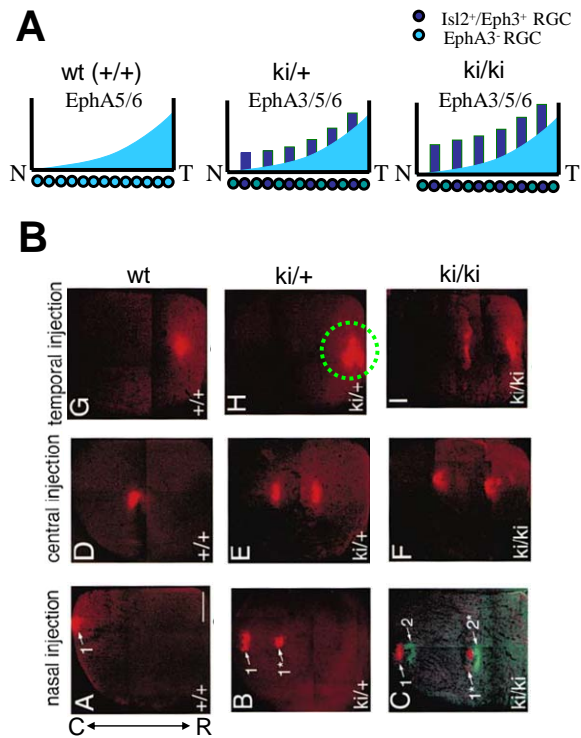

**Figure 3.** Topographic map in Isl2/EphA3 knockin mice. (A) Distribution of the EphA receptors in the retina of wild-type (left), heterozygous (center), and homozygous (right) knockins. (B) Results of anterograde tracing of axons after temporal, central, and nasal injections for these three animals adopted from (Brown et al., 2000). In homozygous knockins (ki/ki, right column) the map is doubled for all three retinal locations. In heterozygous knockins (ki/+, central column) the map is doubled after nasal and central injection, but is single-valued after temporal injection (dashed circle).

Maps in Isl2/EphA homo- and heterozygote mutants differ qualitatively in temporal retina (rostral SC). The amount of additional EphA3 receptor is smaller in heterorozygous than in homozygous knockins by about a factor of two (Brown et al., 2000) (Figure 3A). This fact is reflected in a smaller separation between termination zones of the two types of axons (EphA+ and -) in heterozygous knockins (Figure 3B, central column) compared to the homozygous case (Figure 3B, right column). When temporal axons are traced in heterozygotes however, two populations of axons blend completely forming a single termination zone (green circle in Figure 3). Thus, although chemical labels favor separation between two classes of axons having differing levels of EphA receptor, some additional factors compete with the effects of chemical labels. These additional factors in effect restore the correct topographic quality of the projection from temporal retina despite the disruption of chemical labels. We then investigated computationally these additional factors that could lead to the merging of two groups of axons (EphA3+ and -) in temporal retinas of heterozygous knockins.

## Results of the computational model

Although EphA+ and EphA- axons carry different chemical labels that favor their separation in the target, some additional information about the axon retinal origin is still available. This information is represented in the electric activity of these axons that is correlated between cells neighboring in retina. Correlated neural activity could potentially restore the topographic order in temporal retina resulting in a single-valued map.

Correlated activity enters our model through Hebbian learning rules leading to an effective attraction between axons with correlated activity in the target. Thus, the two classes of axons with different levels of chemical labels (EphA3+ and -) are attracted to each other by the Hebbian mechanisms because they originate from similar locations in the retina, which creates a potential for their collapse into a single TZ. When the difference in the levels of chemical labels is reduced, such as in the heterozygous knock-ins (Figure 4C, central column), the activity-dependent factors restore topographic order.

The effects of chemical labels are the weakest in temporal retina (Figure 5) which leads to the bifurcation in temporal rather than in the nasal retina. This is because the gradient of endogenous EphA is maximal in temporal retina (Feldheim et al., 2000) which leads to the weakest effect of exogenous EphA3 there (Koulakov and Tsigankov, 2004). The attraction between EphA3+ and EphA3- axons has the best opportunity to overcome the chemical factors in the temporal retina. Our model therefore reproduces the collapse of the two branches of topographic map into a single termination zone observed in heterozygous Isl2/EphA3 knock-in mice.

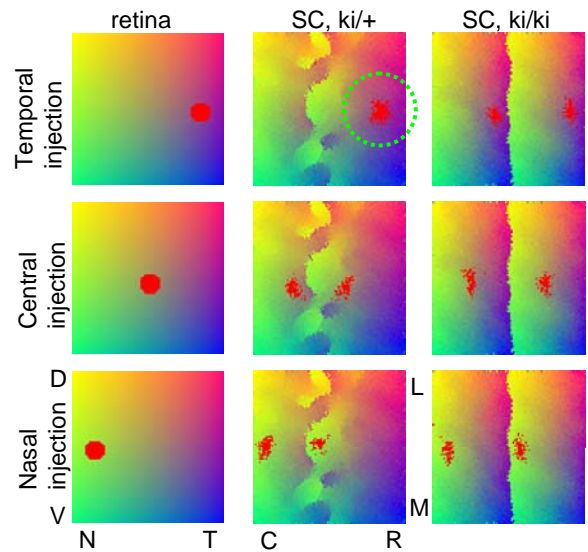

**Figure 4.** The mapping in knock-ins as obtained in the computational model. The complete map structure is color-coded according to the axonal origin in retina as shown by the left column. In addition the termination zones for a small subset of axons indicated by the red points are shown for temporal (upper row), central (center row) and nasal (lower row) labeling. The heterozygous knock-ins (central column) has less amount of the additional EphA receptor than homozygotes (right column). This is to model the results of anterograde labeling as in Figure 3. Comparison with experimental results in Figure 3 shows similarity.

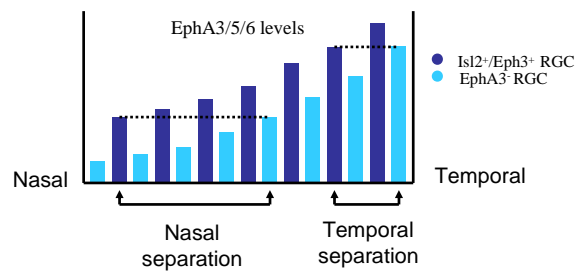

**Figure 5.** The separation between two sets of axons [EphA3+ (blue) and EphA3- (cyan)] is smaller if they originate from temporal retina. Because the gradient of the endogenous EphA receptor is maximal in the temporal retina EphA3- axons in this region have the smallest distance to the EphA3+ axons with similar overall levels of the receptor (dashed lines).

## Difference between transitions induced by correlated activity and noise

We argued before (Koulakov and Tsigankov, 2004) that collapse of two branches of the map in temporal retina of heterozygous Isl2/EphA3 knock-in mice may be induced by stochasticity and noise that limits map's precision. We have showed that if the separation between average locations of EphA3+ and EphA3- axons is smaller than the precision of the map, two types of projections may collapse into one. The finite precision of projections may be due to noise induced by the stochastic nature of axon and dendrite branching (Alsina et al., 2001). Here we argue that an additional factor leading to the collapse of two branches of the map is the effective attraction due to Hebbian mechanisms. Below we analyze the differences between two types of collapse, driven by noise and by correlated activity.

If collapse between two maps is driven by correlated activity (Figure 6A and B), the transition between doubled and single-valued maps is discontinuous. This implies that as the point of transition (Figure 6A point 3) is approached from the nasal direction one of the sets of axons (EphA3- i.e. WT) disappears at one location in the target and moves to a different location to join another set of axons (EphA3+). In the noise-driven case (Figure 6C and D) the collapse is continuous and consists in blending of two broad distributions corresponding to EphA3+ and EphA3- axons as temporal regions of the retina are approached (Figure 6D, 1-4). As a result two distributions of axons do not truly blend in the noise-driven case. Indeed the EphA3- axons (cyan) in Figure 6C are always above EphA3+ axons (blue). This is in contrast to the correlated activity-driven case (Figure 6A), in which these two types of axons are truly mixed in the target if they originate from temporal retina. These features suggest that the transition that is driven by correlated activity is qualitatively different from the noise-driven collapse (Koulakov and Tsigankov, 2004) in a way that can be distinguished experimentally.

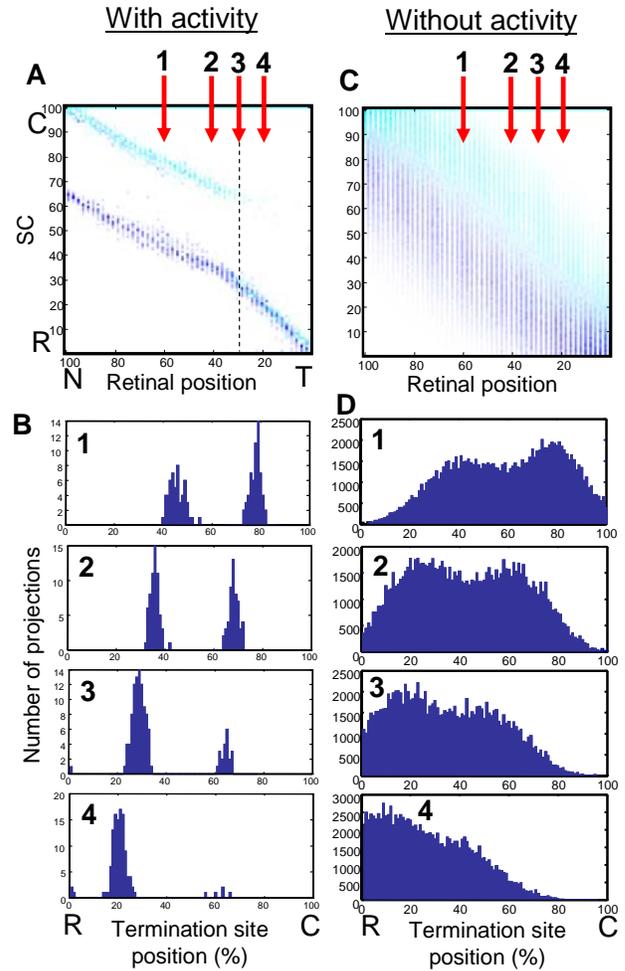

**Figure 6.** Activity-driven collapse (A and B) of two types of projections (EphA3+ and EphA3-) is different from noise-driven collapse (C and D) ($E_{act} = 0$). (A and C) The axonal positions in the target are shown as a function of the postion of their origin in retina. Individual EphA3+/EphA3- axons are represented by blue/cyan points. (B and D) The distribution of terminal positions in the target for subset of axons originating from different retinal locations marked 1-4 in A and C.

## Experimental predictions

The following features of the projection in heterozygous Isl2/EphA3 mutants could be used to confirm that the collapse of EphA3+ and EphA3- braches of the map is indeed driven by Hebbian factors. First, transition is spatially discontinuous. This implies that one projection (EphA3-) disappears as the location of retinal injection is moving from the nasal to the temporal direction. At the same time the disappearing termination site maintains a finite separation from the other branch of the map. This is in contrast to another possibility in which the two branches of the map merge continuously as the injection point is shifted temporarily. Second, it is the wild-type axons (EphA3-) that are actually making the transition. As evident from Figure 6A, the EphA3+ axons (blue) are shifting their locations smoothly as injection point is moved nasally past the collapse point. On the other hand the wild-type axons (cyan) are jumping a finite distance to join the EphA3+ group. Third, the termination site corresponding to the wild-type axons that is located more caudally in Figures 6B is gradually losing its strength as the point of collapse is approached. These three features of the transition are quite specific to the activity-driven collapse of the two maps and could be used to distinguish it from the noise-driven collapse experimentally.

# Discussion

In this study we investigated the interplay between effects of correlated activity and chemospecificity in the development of neural connectivity. We studied the model in which mechanisms based on the chemospecificity and correlated activity can operate independently and thus the contributions due to these two factors are combined additively [equation (1)]. We have shown previously that this form of interaction is sufficient to explain mapping in ephrin-A knockout mice (Tsigankov and Koulakov, 2006). Because correlations in the neural activity could be modulated by external stimuli our model can combine the effects genes and environment in the formation of neural connectivity.

We studied the maps in Isl2/EphA3 knock-in mice. In homozygous Isl2/EphA3 knock-ins the topographic map is fully doubled. This implies that an anterograde injection into any location in retina leads to two termination sites marked in SC (Figure 3B, right column). This form of connectivity can be explained using simple sorting of axons in the target on the basis of their overall expression level of EphA receptors (Koulakov and Tsigankov, 2004). These findings are reproduced by our model (Figure 4, right column). An interesting feature is observed in mapping of Isl2/EphA3 heterozygous knock-ins. In these mice, chemical labels alone should yield doubled map at each retinal location similar to the homozygous case (Koulakov and Tsigankov, 2004). This is because if chemical labels lead to the sorting of axons on the basis of their receptor level then heterozygous maps should differ from homozygous only in the extent of separation between two branches of the mapping. What is observed however is that the map is doubled in nasal retina as predicted by chemical labels but is single valued in temporal retina (Figure 3B, central column). The topographic map therefore experiences bifurcation when going from temporal to nasal retina. In other words, two branches of the doubled map collapse into a single-valued map when the observation point is moved temporarily. Our study centers on the collapse (or bifurcation) that is observed in the maps of heterozygous knock-ins.

We argue that this feature is important because it may occur due to the interplay between the effects of correlated neural activity (Hebb) and chemical labels (Sperry). Indeed the chemical labels, such as Eph receptors and ephrins, favor doubled map throughout the knock-in retina. At the same time correlated neural activity favors single-valued mapping. This is because correlated activity carries additional information about an axonal geometric origin in retina that is independent on chemical labels. This information leads to an effective attraction between axons with correlated activity that originate from neighboring points in retina. As a result activity-dependent attraction may overcome the separation induced by chemical labels and two branches of the map coalesce in temporal retina (Figure 6A), where the effects of chemical labels are the smallest (Figure 5). The point of bifurcation (Figure 6A, dashed line) is established from the balance of costs due to chemical labels (Sperry) and correlated activity (Hebb), as elaborated in Methods. We suggest experimental tests that could validate this scenario.

## Alternative models for retinocollicular connectivity formation.

The most prominent alternative models include the dual gradient model (DGM) (Gierer, 1981; Yates et al., 2004) and the servomechanism model (SVM) (Fraser and Perkel, 1990; Honda, 1998, 2003). The former assumes that axons are influenced by two gradients in the target. In the simplest form axons are guided by gradients of two repellents: one having maximum in the rostral while the second having maximum in the caudal

SC. The correct location for a termination zone for an axon is established through balancing the repulsive forces from these gradients. The second class of models represented by SVM assumes that axons terminate at locations with particular level of ligand corresponding to the expression level of receptor on the axon. This model was suggested recently to explain the mapping along the vertical dorso-ventral axis in the retina. Both of these models postulate the matching principle whereby the temporal axons will have preference to the rostral SC and will be repelled from the caudal SC. Similarly nasal axons preferentially grow in the caudal SC while are repelled from rostral SC in both DGM and SVM mechanisms. The latter prediction of the matching principle models contradicts to the striped assays experiment in which nasal axons, when exposed to alternating stripes extracted from caudal and rostral SC do not show any preference in their extension (Walter et al., 1987a; Walter et al., 1987b). This observation is difficult to explain by the presence of competition between axons *in vitro*, because temporal axons do demonstrate preferential growth in the rostral stripes in the same experiment. In addition, both DGM and SVM require competition between axons *in vivo* to explain topographic map compression/expansion observed after tectal/retinal lesions (Fraser and Perkel, 1990).

The model that we employ in this study is consistent with the striped assay experiments. Indeed, nasal axons express low levels of EphA receptor and, therefore, are unresponsive to the difference between rostal and caudal stripes as observed in the striped assay experiments (Walter et al., 1987a). Temporal axons on the other hand have high levels of EphA receptor and, therefore, are repelled by the caudal stripes. This is because the latter have high levels of expression of the ephrin-A repellents (Figure 1). We thus adopt the model based on single repellent and competition (Tsigankov and Koulakov, 2006) because it is both parsimonious and consistent with the striped assay experiments.

**Mapping in Isl2/EphA3+/EphA4-**

In recent studies the retinocollicular projection was studied in mice in which the anomalous distribution of EphA3 receptor is combined with the decrease in the expression of EphA4 (Reber et al., 2004). The latter is expressed at a constant level throughout RGC layer. Topographic projections in these animals are similar qualitatively to the Isl2/EphA3 case, i.e. maps are double-valued in homozygous and bifurcate in heterozygous knock-ins. The point of bifurcation however is shifted in these mutants. Another set of qualitative differences hints at the active role that competition between axons for space may play in these animals. Thus, the EphhA3+ branch of the map is substantially compressed occupying in some cases less than 30 percent of the target. In many instances the EphA3+ and EphA3- branches of the map are separated by a gap. The exact nature of the axons projecting to this gap is not clear. Our present model does not allow addressing these features since the area occupied by each axon is fixed in our study. In our model the reduction in the level of EphA receptor, such as in EphA4- mutants, leads to results similar to the ones presented above. Results in (Reber et al., 2004) suggest more complex interactions of the molecular labels with the effects of competition than those used in our study. We argue however that qualitative features of the bifurcation driven by activity can be understood on the basis of our model. A more refined model of interaction between molecular labels and competition could be advanced when mapping in Isl2/EphA3+/EphA4- mice is more completely understood.

**Other models for bifurcation in heterozygote mutants**

(Honda, 2003) put forward a model which obtains the double-valued maps in homozygote knock-ins. However the bifurcation of the maps in heterozygous case is not reproduced. Other models (Koulakov and Tsigankov, 2004; Yates et al., 2004; Willshaw, 2006) yield the results resembling those is Figure 6C. These models therefore suggest an explanation to map's doubling (bifurcation) observed in heterozygous knock-ins. In these studies EphA3+ and EphA3- projections, although blended, are not truly merged in temporal retina. This implies that there is a bias for EphA3+/EphA3- axons to terminate more rostrally/caudally despite proximity of these projections. Here we suggest a qualitatively different solution in which two types of projections form the same termination site with no noticeable bias (Figure 6A). We provide several qualitative features that could distinguish the activity-dependent mechanism proposed here from the earlier models.

## Methods

### Description of the model

Our model is designed to predict the locations of terminations of retinal axons. More exactly, our model reproduces the behavior of synapses formed by axons. The synapses are defined by the weight matrix $W_{ij}$, where the index $j$ describes the number of retinal axon, while the index $i$ is the number of the dendrite with which given synapse is formed. The weight matrix therefore describes the strength of connection between the axon number $j$ and dendrite number $i$. We then define the affinity energy that is a function of the weight matrix. The affinity energy is a sum of the chemoaffinity (Sperry) and correlated activity-dependent (Hebb) contributions, as postulated by equation (1). The affinity energy is similar to the one used by us before (Koulakov and Tsigankov, 2004; Tsigankov and Koulakov, 2006) with the expression levels of chemical labels modified to address experiments in Isl2/EphA3 mutants (Brown et al., 2000). The Sperry contribution is

$$E_{chem} = \sum_{\alpha\beta} M_{\alpha\beta} \sum_{ij} L_i^\alpha W_{ij} R_j^\beta \qquad (2)$$

Here indexes $\alpha$ and $\beta$ describe the chemical labels ($\alpha =$ EphA or B, $\beta =$ ephrin-A or -B), while the matrix $M_{\alpha\beta}$ defines the affinities between pairs of labels, such as receptor and a ligand. The parameters $R_j^\beta$ and $L_i^\alpha$ are the receptor and ligand expression levels of axon number $i$ and dendrite number $j$ respectively. The Hebb contribution

$$E_{act} = -\frac{1}{2} \sum_{ijml} C_{ij} W_{mi} W_{lj} U_{ml} \qquad (3)$$

Here $C_{ij}$ is the correlation in electric activity between axons number $i$ and $j$. This function describes the strength of similarity between axons as a function of their location is retina. The arbor function $U_{ml}$ on the other hand describes the strength of Hebbian interaction as a function of dendrite position in the target. The affinity energy defined by equations (1) through (3) is minimized using the stochastic procedure defined below.

RGCs are arranged in retina on a square N by N array with N=100. These array of cells establish connections with a matching in size square array of collicular dendrites. Each axon is constrained to make connections with one and only one collicular dendrite. We therefore assume that $W_{mi} = 1$ for the pair of cells $m$ and $l$ that are connected and 0 for unconnected cells. This assumption implements the competition constraint described in the text.

We begin from a random set of connections that reflects the broad initial distribution of axons and their synapses in the target (O'Leary and McLaughlin, 2005). To minimize the affinity energy (1) we use an iterative stochastic optimization procedure. On each step of the algorithm two cells are chosen randomly. The cells are not necessarily neighboring in the target or in the retina. We then calculate the potential change in the affinity energy for the modification of retinocollicular connectivity in which these two cells exchange their positions in the target. This change in energy is defined by $\Delta E$. The modification of connectivity is then implemented with probability

$$p_{exchange} = \frac{1}{1 + \exp(4\Delta E)} \qquad (4)$$

Thus, if the energy is decreased as the result of this modification ($\Delta E < 0$) the probability to accept this attempt is more than 1/2, leading therefore to the bias towards minimizing the overall value of energy. This step is repeated $10^7$ times. The number of iteration is chosen to ensure the algorithm's convergence for the wild type distribution of the molecular labels.

The parameters of the model are as follows. The distributions of molecular labels are

$$R_{EphA} = \exp(-x/N) - \exp([x-2N]/N) + \Delta R \qquad (5)$$

$$L_{ephrin-A} = \exp([x'-N]/N) - \exp(-[x'-N]/N) \qquad (6)$$

$$R_{\text{EphB}} = \exp(-y/N) \quad (7)$$

$$L_{\text{ephrin-A}} = \exp(-y'/N) \quad (8)$$

Here the horizontal and vertical coordinates in retina are $x$ and $y$, while the collicular coordinates are denoted by $x'$ and $y'$. All coordinates vary between 1 and $N$. The distribution of EphA and ephrin-A receive negative corrections in equations (5) and (6) due to masking by ephrin-A and EphA present in retina and SC respectively, as discussed in (Tsigankov and Koulakov, 2006). The additional level of expression of EphA receptor $\Delta R$ is equal to 0.35 and 0.7 for heterozygous and homozygous conditions for axons with even $x$ only and is zero otherwise.

The matrix of affinities $M_{\alpha\beta}$ in equation (2) is

$$M_{\text{EphA,ephrin-A}} = -M_{\text{EphB,ephrin-B}} = 30 \quad (9)$$

$$M_{\text{EphA,ephrin-B}} = M_{\text{EphA,ephrin-B}} = 0 \quad (10)$$

Negative/positive values of the matrix of affinities describe chemoattraction/repulsion. The zero values imply that there is no direct interaction between the "A" and "B" families of receptors and ligands (Flanagan and Vanderhaeghen, 1998). The parameters in equation (3) are as follows

$$C_{ij} = \exp\left(-|\vec{r}_i - \vec{r}_j|/a\right) \quad \mathbf{(11)}$$

$$U_{ml} = \gamma \exp\left([\vec{r}_m - \vec{r}_l]^2 / 2b^2\right) \quad \mathbf{(12)}$$

where $a = 0.11N$ is the range of correlations in the retina (McLaughlin et al., 2003b; Tsigankov and Koulakov, 2006) while $b = 0.03N$ and $\gamma = 0.25$ are the range and the strength of Hebbian attraction in SC (Tsigankov and Koulakov, 2006).

**Position of bifurcation**

Here we calculate the location of the bifurcation point from the balance between Hebbian and Sperry contribution to the affinity energy. The bifurcation in the map is associated with the interface between the single-valued and doubled maps. In the doubled map the Sperry contribution to the affinity functional is minimized, while the Hebbian contribution is increased by

$$\Delta E_{act} = \int_0^x n d\tilde{x} \int_0^N dy\, U_H \quad (13)$$

Here $n$ is the density of neurons ($n=1$ in our model), $x$ is the location of the interface, $U_H \sim \gamma n b^2$ is the Hebbian energy per neuron. The Hebbian energy is estimated here up to the numerical factor which depends of the exact geometry of the problem. In the single-valued part of the map the Hebbian contribution has the minimum possible value while the Sperry contribution is increased. The total increase in the Sperry contribution is

$$\Delta E_{chem} = \int_x^N n d\tilde{x} \int_0^N dy\, U_{Sp}\,. \quad (14)$$

Here $U_{Sp}$ is the Sperry contribution per neuron. The total affinity is minimum if

$$\frac{dE}{dx} = \frac{d\Delta E_{chem}}{dx} + \frac{d\Delta E_{act}}{dx} = 0 \quad (15)$$

Due to equations (13) and (14) this implies that

$$U_H = U_{Sp} \quad (16)$$

To evaluate the increase in the Sperry contribution in the area occupied by the single-valued contribution we notice that it is equal to

$$U_{Sp} = M_{\text{EphA,ephrin-A}} \cdot \nabla R \nabla L \cdot \Lambda^2 \quad (17)$$

Here $\Lambda$ is the shift of the axons in the single-valued map from the location minimizing Sperry contribution. This shift is therefore equal to the separation between two branches of the doubled map. The gradients of the wild-type levels of receptor and ligand are denoted by $\nabla R$ and $\nabla L$. Because this correction to energy per neuron describes deviation from the minimum it is quadratic in $\Lambda$. The assumption under which (17) is true is that $\Lambda \ll N$ i.e. maps separation in the doubled map is smaller than the size of the map. To find $\Lambda$ we notice that for doubled map $R(x + \Lambda) = R(x) + \Delta R$. This implies that the wild-type EphA3- axons with the level of EphA expression $R(x + \Lambda)$ terminate at the same location as the knockin EphA3+ axons with the receptor levels of $R(x) + \Delta R$. Thus the separation between two maps is

$$\Lambda = \frac{\Delta R}{\nabla R} \qquad (18)$$

Combining (16), (17), and (18) we obtain the equation for the location of the point of doubling $x$

$$\frac{\nabla L(x)}{\nabla R(x)} = \frac{U_H}{M_{\text{EphA,ephrin-A}} \Delta R^2} \qquad (19)$$

The single-valued are of the map is defined by the condition

$$\frac{\nabla L(x)}{\nabla R(x)} < \frac{U_H}{M_{\text{EphA,ephrin-A}} \Delta R^2}, \qquad (20)$$

which implies that the Hebb contribution is large. The doubled map region is defined by the opposite to equation (20) condition. This confirms the qualitative understanding that we derived from Figure 5 that collapse should occur more readily at location where the gradient of ligand is small and gradient of wild-type level of receptor is large i.e. in temporal retina, as observed experimentally (Figure 3).